\documentclass{webofc}
\usepackage{ulem}
\usepackage[varg]{txfonts}
\usepackage{hyperref}
\usepackage{cleveref}
\usepackage[most]{tcolorbox}
\usepackage{url}
\hypersetup{colorlinks=true,citecolor=blue,urlcolor=blue,linkcolor=blue}
\RequirePackage{xspace}

\usepackage{listings}
\usepackage{listings}
\usepackage{xcolor}

\definecolor{codegreen}{rgb}{0,0.6,0}
\definecolor{codegray}{rgb}{0.5,0.5,0.5}
\definecolor{codepurple}{rgb}{0.58,0,0.82}
\definecolor{backcolour}{rgb}{0.95,0.95,0.92}

\definecolor{codepurple}{rgb}{0.58,0,0.82}
\definecolor{backcolour}{rgb}{0.95,0.95,0.92}
\definecolor{keywordcolor}{rgb}{0.36,0.54,0.66}
\definecolor{commentcolor}{rgb}{0.4,0.4,0.4}

\lstdefinelanguage{yaml}{
  morekeywords={true,false,null,yes,no},
  sensitive=true,
  morecomment=[l]{\#},
  morestring=[b]',
  morestring=[b]",
  alsoletter={:},
  keywordstyle=\color{keywordcolor}\bfseries,
  commentstyle=\color{commentcolor}\itshape,
  stringstyle=\color{codepurple},
}

\lstdefinestyle{yamlstyle}{
  backgroundcolor=\color{backcolour},
  basicstyle=\ttfamily\small,
  breaklines=true,
  showstringspaces=false,
  columns=fullflexible
}

\lstdefinestyle{pythonstyle}{
    backgroundcolor=\color{backcolour},   
    commentstyle=\color{codegreen},
    keywordstyle=\color{magenta},
    numberstyle=\tiny\color{codegray},
    stringstyle=\color{codepurple},
    basicstyle=\ttfamily\footnotesize,
    breakatwhitespace=false,         
    breaklines=true,                 
    captionpos=b,                    
    keepspaces=true,                 
    numbers=left,                    
    numbersep=5pt,                  
    showspaces=false,                
    showstringspaces=false,
    showtabs=false,                  
    tabsize=2
}

\lstdefinestyle{plain}{
  basicstyle=\ttfamily \footnotesize,     
  keywordstyle=,
  commentstyle=,
  stringstyle=,
  showstringspaces=false,
  breaklines=true,
  language=,
}

\crefname{flarebox}{Listing}{Listings}
\Crefname{flarebox}{Listing}{Listings}
\crefname{figure}{Fig.}{Figs.}
\Crefname{figure}{Fig.}{Figs.}
\definecolor{codeboxbg}{rgb}{0.95,0.98,0.96}
\definecolor{codeboxframe}{rgb}{0.60,0.78,0.86}
\newtcblisting[use counter=flarebox]{codebox}[2][]{%
  breakable,
  colback=codeboxbg, colframe=codeboxframe,
  boxrule=0.5pt, arc=1pt,
  left=3pt, right=3pt, top=3pt, bottom=3pt,
  fonttitle=\footnotesize, coltitle=black,
  title={Listing~\thetcbcounter.~#2},
  label={#1},
  listing only,
  listing options={style=yamlstyle, language=yaml, backgroundcolor=\color{codeboxbg},
    basicstyle=\ttfamily\fontsize{7.65pt}{9.35pt}\selectfont\color{black}, keywordstyle=\color{black},
    commentstyle=\color{black}, stringstyle=\color{black}, numbers=none},
}
\newtcolorbox[use counter=flarebox]{cmdbox}[2][]{%
  breakable,
  colback=codeboxbg, colframe=codeboxframe,
  boxrule=0.5pt, arc=1pt,
  left=3pt, right=3pt, top=3pt, bottom=3pt,
  fonttitle=\footnotesize, coltitle=black,
  fontupper=\ttfamily\fontsize{7.65pt}{9.35pt}\selectfont,
  title={Listing~\thetcbcounter.~#2},
  label={#1},
}

\def\key{\textsc{Key4HEP}\xspace}
\def\flare{\textsc{Flare}\xspace}
\def\btwoluigi{\textsc{b2luigi}\xspace}

\def\whizard{\textsc{Whizard}\xspace}
\def\madgraph{\textsc{MadGraph5}\xspace}
\def\pythiaeight{\textsc{Pythia8}\xspace}
\def\pythiasix{\textsc{Pythia6}\xspace}
\def\delphes{\textsc{Delphes}\xspace}
\def\fccanalyses{\textsc{FCCAnalyses}\xspace}
\def\htcondor{\textsc{HTCondor}\xspace}
\def\slurmsys{\textsc{Slurm}\xspace}
\def\lsf{\textsc{LSF}\xspace}
\def\pydantic{\textsc{Pydantic}\xspace}
\def\lhereader{\textsc{LHEReader}\xspace}
\def\rootsw{\textsc{ROOT}\xspace}

\begin{document}

\title{\textsc{Flare}: an open-source data workflow orchestration tool}

\author{
\firstname{Cameron Cooper} \lastname{Harris}\inst{1}\fnsep\thanks{Speaker, \email{cameron.harris@adelaide.edu.au}} \and
\firstname{Aman} \lastname{Desai}\inst{1}\fnsep\thanks{\email{aman.desai@adelaide.edu.au}} \and
\firstname{Paul} \lastname{Jackson}\inst{1}\fnsep\thanks{\email{p.jackson@adelaide.edu.au}}
}

\institute{Department of Physics, Adelaide University, Adelaide, SA 5005, Australia}

\abstract{
\textsc{Flare} is an open-source Python-based data workflow orchestration tool powered by \btwoluigi. It automates the workflow of Monte Carlo (MC) generators inside the \key stack, such as \textsc{Whizard}, \textsc{MadGraph5} and \textsc{Pythia8}, together with fast detector simulation using \textsc{Delphes}. It also automates the Future Circular Collider (FCC) physics analysis software workflow. These workflows are combined, giving a user an automated pipeline from MC production to final \textsc{FCCAnalyses} histograms. With its customisation options and Python API, \textsc{Flare} simplifies executing FCC-ee analyses, especially when custom MC events are required.
}

\maketitle


\section{Introduction}\label{intro}

The \key stack~\cite{Key4hep:2023nmr} is a software solution that gathers HEP software for future high-energy particle colliders, such as the Future Circular Collider (FCC), into a single place, spanning Monte Carlo (MC) event generation, parametric and full simulation, and analysis tools such as \fccanalyses~\cite{helsens_2025_15528870}, the common framework for FCC-ee physics analyses~\cite{FCC:2025lpp}. This package is open source and available through the CernVM-File System (CVMFS)~\cite{Blomer:2011zz}.

\begin{figure}[htb]
    \centering
    \includegraphics[width=\linewidth]{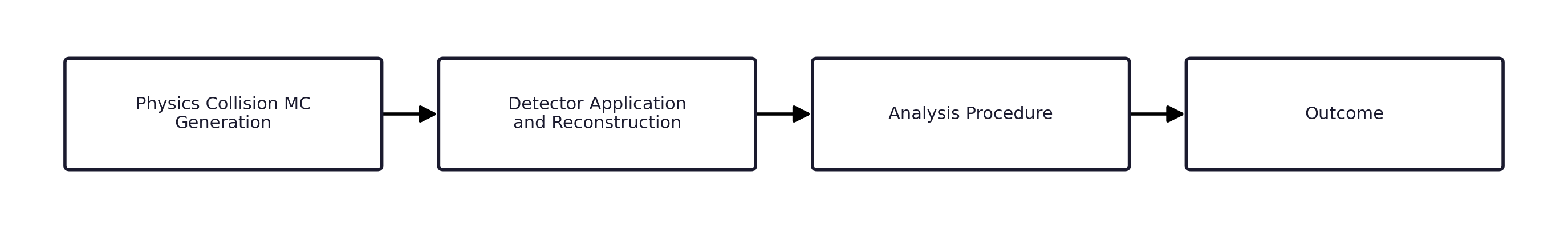}
    \caption{The basic workflow for HEP MC production and analysis. It begins with the physics collision events to study, e.g.\ $e^+e^- \to ZH$. The user applies a detector configuration and a reconstruction technique (fast or full simulation), and finally an analysis procedure runs on the reconstructed events.}
    \label{fig:basic-key4hep-study-workflow}
\end{figure}

Automating the coordinated execution of HEP software is crucial for a seamless, reproducible physics MC production and analysis pipeline. We introduced \flare~\cite{CooperHarris:2025lqd} (FCCee b2Luigi Automated Reconstruction and Event processing) to address this: it runs a physics analysis from MC generation through to final results such as histograms or plots (\Cref{fig:basic-key4hep-study-workflow}), relying on the HEP software shipped with \key.

\flare is an open-source workflow orchestration tool built using \btwoluigi~\cite{b2luigi2025, luigi2025}, the workflow management tool maintained by the Belle II Collaboration. \btwoluigi has a Pythonic API based on classes known as \textit{Tasks}, each containing executable code that produces an output on the local system. Multiple Tasks are joined into a Directed Acyclic Graph (DAG) that defines the order in which they run. A built-in scheduler identifies incomplete Tasks whose upstream dependencies are all complete and executes them; this process, orchestration, manages a workflow from start to finish. \btwoluigi is the bedrock of \flare, letting us package \key tools into Tasks and execute them in the required order. It also provides a ``one-line'' procedure for submitting to batch systems such as \htcondor~\cite{condor-practice}, \slurmsys~\cite{slurm} and \lsf~\cite{IDM-lsf}, which \flare leverages for submission and execution.

Previously, \flare was presented at the DRD4 general meeting\footnote{\tiny\url{https://indico.cern.ch/event/1473150/timetable/?view=standard\#12-flare-an-open-source-data-w}} and PyHEP 2025\footnote{\tiny\url{https://indico.cern.ch/event/1566263/timetable/?view=standard\#2-flare-fccee-b2luigi-automate}} in October 2025, and at the FCC Munich Workshop\footnote{\tiny\url{https://indico.cern.ch/event/1588696/timetable/?view=standard}} in January 2026, when it was integrated into the \key stack and made available via CVMFS.

Developments since its inclusion in \key include, among others, more MC generators and new ways to customise an MC generation workflow with different MC generator and reconstruction tool combinations; a command-line interface (CLI) linting tool that checks all input files for formatting and compatibility before runtime; and the \texttt{Add Stage} feature, which lets a user bundle custom code into its own \flare Task, choose where in the DAG it is added, and rely on \flare to pipe all inputs into it correctly. Development of the full-simulation workflow is in progress.

\section{\textsc{Flare} Software}\label{sub:flare-software}

The primary idea behind the \flare package was to build a framework enabling users to create and execute automated and complex workflows with standard \key tools, through a simplified user interface.

It was designed with the following principles:
\begin{enumerate}
    \item \flare has full control and will manage all inputs and outputs of any given Task.
    \item For each Task that requires a user's input file, there must exist a unique file.
\end{enumerate}

The first principle ensures that the software can orchestrate and execute a workflow, with all generated files located correctly in the local file system. The second principle is that, before a workflow runs, \flare verifies that every input file from the user is present in the working directory. To ensure all Tasks in the DAG are verified, each Task requiring a user's input file must be unique.

The \flare implementation relies on an interface build pattern where the interface is the set of input YAML files, whose data structures are validated with \pydantic~\cite{pydantic}. For each \key tool, there exists a Workflow in the form of a YAML file. The Workflow defines the individual Tasks executed for a given software tool. \Cref{fig:flare-workflow-example} shows an example Workflow for MC production using \madgraph~\cite{Alwall_2014} followed by parton shower and fast simulation of the \madgraph output. The Workflow is made up of three individual Tasks with a structure inspired by GitHub Actions. The order of Tasks is inferred by their declaration order inside the YAML and is preserved at runtime. The leading Task is defined under the header Stage 1 and includes three required fields: \texttt{cmd}, \texttt{args} and \texttt{output\_file}. The \texttt{cmd} defines the executable subprocess command for a given Task. The \texttt{args} defines the input arguments required by \texttt{cmd}. The \texttt{output\_file} represents the name of the output file produced upon Task completion. The \madgraph implementation of \Cref{fig:flare-workflow-example} has three separate Tasks. Stage 1 shows the \madgraph tool execution and Stage 2 is the fast-simulation reconstruction using the output from Stage 1 along with the user-defined input files shown under \texttt{args}.

\begin{codebox}[fig:flare-workflow-example]{\madgraph implementation for a \flare Workflow. The data structure used to create Tasks and establish their order.}
madgraph:
    stage1:
        cmd: mg5_aMC {0}
        args:
            - ++_runcard.dat
        on_completion:
            - madgraph_move_contents_to_tmp_output
        output_file: signal.lhe
    stage2:
        cmd: DelphesPythia8_EDM4HEP {0} {1} {2} {3}
        args:
            - card_<>.tcl
            - edm4hep_<>.tcl
            - pythia_card_<>.cmd
            - ().root
        output_file: output_.root
        pre_run:
            - madgraph_copy_lhe_file_to_cwd
\end{codebox}

\flare can connect different workflows to build complex Task graphs that would otherwise be challenging to manage.

The \flare CLI tool is the interface the user interacts with to run their workflows. It is accessible by (i) sourcing \key from CVMFS, or (ii) installing the PyPI package (\Cref{fig:pip-install}).

\begin{cmdbox}[fig:pip-install]{Terminal command to install \flare from the PyPI package manager}
\$ pip3 install hep-flare
\end{cmdbox}

An \fccanalyses Workflow can be executed with \flare, assuming all input files are in the working directory and a \texttt{flare.yaml} is defined. A \texttt{flare.yaml} file instructs \flare to use a specific \texttt{batch\_system} for submitting computational jobs. This \texttt{batch\_system} variable is connected directly with \btwoluigi, which is the underlying software that handles the submission and monitoring of all batch jobs. The command shown in \Cref{fig:fccanalyses-cli-command} will run the Workflow for \fccanalyses.

\begin{cmdbox}[fig:fccanalyses-cli-command]{CLI command to run the \fccanalyses workflow}
\$ flare run analysis
\end{cmdbox}

The MC production tools available in \flare are \madgraph~\cite{Alwall_2014}, \whizard~\cite{Kilian_2011}, \pythiasix~\cite{Sjostrand:2006za} and \pythiaeight~\cite{Sjostrand:2014zea}. At present, only fast simulation, offered by \delphes~\cite{deFavereau:2013fsa}, is integrated into \flare. The CLI command to execute is shown in \Cref{fig:mcprod-cli-command}.

\begin{cmdbox}[fig:mcprod-cli-command]{CLI command to run the MC Production workflow}
\$ flare run mcproduction
\end{cmdbox}

Consider the standard HEP scenario in which a user wishes to produce MC samples and then analyse them with an analysis package (for example \fccanalyses). As before, all input files must sit in the working directory. To run the workflow, the user can execute the command shown in \Cref{fig:mcprod-analysis-cli-command}.

\begin{cmdbox}[fig:mcprod-analysis-cli-command]{CLI command to run the combined MC Production and \fccanalyses workflows}
\$ flare run analysis --mcprod
\end{cmdbox}

\flare can be used with software distributed with \key as well as additional software available from the user's computing environment. This is achieved by adding the software to the system's \texttt{\$PATH} variable. It therefore extends to many collider types and HEP studies. This lets, for example, a user supply their own \madgraph models and modify the internal workings of the software to enable functionality not available from the fixed \key version of \madgraph.

\section{\textsc{Flare} Examples and Further Details}\label{sec:flare-examples}

The repository with examples is available at \url{https://github.com/CamCoop1/FLARE-examples}.

\subsection*{\textsc{Flare} MC Production on Batch System}\label{subsec:large-mc-prod-examples}

An example of MC generation with \whizard interfaced to \pythiasix and \delphes is presented here, together with submission of the resulting jobs to the \slurmsys batch system. The \texttt{flare.yaml} file (\Cref{fig:flare-yaml}) gives the general \flare settings. Here the \texttt{batch\_system} is configured to \texttt{slurm}. The settings are directly passed on to the \btwoluigi backend, ensuring that each Task in the Workflow is submitted to the \slurmsys batch system.

The working directory is organised such that all MC input files along with the \delphes cards are placed in \texttt{mc\_production} subdirectory. Each \whizard input file has a unique name, following \flare principle 2. \Cref{fig:large-mc-prod-yaml} shows the configuration file for defining the MC production tool as well as the exact MC files to be produced. This setup can be executed with the command in \Cref{fig:mcprod-cli-command}.

\begin{codebox}[fig:flare-yaml]{Example of the \texttt{flare.yaml} file used to declare the general settings. It includes the \texttt{batch\_system} variable, which is passed directly to \btwoluigi to declare which batch system to submit Tasks to.}
# flare.yaml

# FLARE Settings
# output directory becomes data/not_foo
name : CHEP-example

# batch settings
batch_system : slurm
slurm_settings:
  mem: 5GB
\end{codebox}

\begin{codebox}[fig:large-mc-prod-yaml]{MC Production \texttt{flare\_mc.yaml} file example in which a production type along with specific datatypes is defined.}
# mc_production/flare_mc.yaml

# Which generator to use on all datatypes
global_prodtype : whizard

# Datatypes to be generated
# Must match file names in mc_production directory
datatype:
    - wzp6_ee_bbH_HWW_ecm240
    - wzp6_ee_bbH_Hbb_ecm240
    - wzp6_ee_ccH_HWW_ecm240
    - wzp6_ee_eeH_HZZ_ecm240
    - wzp6_ee_mumuH_Hbb_ecm240
    - wzp6_ee_nunuH_Hbb_ecm240
    - wzp6_ee_qqH_HWW_ecm240
    - wzp6_ee_ssH_HWW_ecm240
    - wzp6_ee_ssH_HZZ_ecm240
\end{codebox}

\subsection*{\textsc{Pythia8} and Fast Simulation Features}\label{subsec:pythia8}

\flare can also generate MC events with \pythiaeight interfaced to the \delphes fast-simulation reconstruction software in \key. The fast-simulation reconstruction (\Cref{fig:pythia8-cli}) wraps a given detector configuration (\texttt{<detector card>.tcl}) together with the output of the MC generation step.

Moreover, we highlight that \flare simplifies multi-detector studies since a user need only place any number of uniquely named \texttt{.tcl} files with varying detector parameters inside their \texttt{mc\_production} subdirectory and include them in the \texttt{flare\_mc.yaml}. No further configuration is needed. On executing the run command from \Cref{fig:mcprod-cli-command}, \flare takes all unique combinations of the \texttt{datatype} and \texttt{card} lists and schedules a separate Task for each. An example of this can be found in the examples repository.\footnote{\tiny\url{https://github.com/CamCoop1/FLARE-examples/blob/main/MCProduction\_workflow/multi\_detector\_card\_whizard\_example/mc\_production/flare\_mc.yaml}}

\begin{cmdbox}[fig:pythia8-cli]{\delphes Fast Simulation CLI tool execution pattern}
\$ DelphesSTDHEP\_EDM4HEP \textcolor{blue}{<detector card>.tcl} edm4hep.tcl \textcolor{red}{<MC type config>.cmd} output.root
\end{cmdbox}

\subsection*{Outlook for Full Simulation Workflow in \textsc{Flare}}\label{subsec:full-sim}

Full simulation within \flare, implemented via the \key \texttt{k4run} reconstruction~\cite{Key4hep:2023nmr}, is currently being explored. At present it is possible to adjust \texttt{global\_prod\_type} to include \texttt{\_fullsim} at the end, to enable full-simulation studies, for example, for a \whizard workflow with full simulation, use \texttt{global\_prod\_type=whizard\_fullsim}. An example is available in the examples repository.\footnote{\tiny\url{https://github.com/CamCoop1/FLARE-examples/tree/fullsim\_examples/MCProduction\_workflow/fullsim\_examples/whizard}}

\subsection*{Phenomenology and FCC-hh Workflow Developments}\label{subsec:pheno-fcchh}

In our preliminary exploration, the following choices were made for phenomenology workflows:

\begin{itemize}
    \item FCC-hh / LHC hadron colliders use a local installation of \madgraph for MC production
    \item The phenomenology study uses the \texttt{Add Stage} feature, with \lhereader~\cite{Desai:2026lst} as an example, to convert Les Houches Event (LHE) files to \rootsw~\cite{Brun:1997pa}.
\end{itemize}

Usage of such local installations is possible because \flare is able to decouple from the \key stack. The local installation can be included by adding the path to the \texttt{\$PATH} variable.

\section{Additional Features}\label{sec:additional-features}
\flare provides two further tools: the \texttt{Add Stage} feature and a CLI linting tool.

\subsection*{\textsc{Flare} Add Stage}\label{subsec:flare-add-stage}

Users often need Tasks beyond the built-in Workflows. This led to the development of the Add Stage feature. This feature allows the user to bundle their scripts or packages into \flare Tasks which can be scheduled and executed automatically. A simplified example is presented here; a more complete example is also available.\footnote{\tiny\url{https://github.com/CamCoop1/FLARE-examples/tree/fullsim\_examples/FCCAnalysis\_workflow}}

The example of \texttt{flare.yaml} configuration is shown in \Cref{fig:flare-add-stage-yaml}. In this example, additional Tasks are added to the existing \fccanalyses Workflow. The Add Stage feature works similarly to the Workflow interface, requiring \texttt{cmd}, \texttt{args} and \texttt{output\_file}. Additionally, a \texttt{requires} field, a \texttt{required\_by} field, or both are defined for a Task. In this way, a user instructs the software exactly where to insert their Task in the DAG.

\begin{codebox}[fig:flare-add-stage-yaml]{This segment of the \texttt{flare.yaml} file shows how a user can use the \texttt{add\_stage} argument to define their own custom \btwoluigi Tasks and add them to the existing \fccanalyses Workflow.}
# flare.yaml
add_stage:
       TrainBDT:
        cmd: python3 {0}
        args:
          - ml_<>.py
        output_file: output.model
        requires: 'Stage2'

       Combine:
        cmd: python3 {0}
        args:
           - combine_<>.py
        output_file: output.model
        requires: 'trainBTD'

       Move:
        cmd: python3 {0}
        args:
           - move_<>.py
        output_file: output.model
        requires: 'stage2'
        required_by : ['plots']
\end{codebox}

\subsection*{\textsc{Flare} Linting Tool}\label{subsec:flare-lint}

The \flare linting tool is designed to scan the input files required by the Workflows, ensuring compliance with the requirements of the software. First, it checks that an \texttt{inputDir} and an \texttt{outputDir} are defined. Second, it checks that any path-like strings in the input files are f-strings that begin with \texttt{inputDir} or \texttt{outputDir}. Users may add their own subdirectories for bookkeeping, but \flare sets \texttt{inputDir} and \texttt{outputDir} itself. At present, the only Workflow that requires this linting treatment is the \fccanalyses Workflow. The linting tool is executed with the command in \Cref{fig:flare-lint}.

\begin{cmdbox}[fig:flare-lint]{CLI command for linting the \fccanalyses Workflow along with any Add Stage Tasks defined by the user}
\$ flare lint fccanalysis
\end{cmdbox}

\section{Conclusions}\label{conclusions}
\flare is a workflow orchestration tool that aims to simplify working with the tools provided by the \key stack. It is powered by \btwoluigi, which orchestrates and runs the Workflows. It can also submit entire Workflows to a user-specified batch system such as \slurmsys or \htcondor. The two main Workflows are the MC Production Workflow, in which tools such as \whizard generate specific collision events and the \delphes fast simulation then reconstructs them, and the \fccanalyses Workflow. \flare is designed such that any Workflow can easily be piped into another with minimal additional input from the user. The \flare CLI tool simplifies executing the Workflows. Development is currently focused on integrating full simulation, alongside exploration of phenomenological and FCC-hh studies.

\section*{Acknowledgements}

We would like to thank Juraj Smiesko, Juan Carceller, the \btwoluigi maintainers Jonas Eppelt and Alexander Heidelbach, and Jonathan Woithe for their help and useful discussions.

\bibliography{biblio}

@article{Kilian_2011,
    author = "Kilian, Wolfgang and Ohl, Thorsten and Reuter, Jurgen",
    title = "{WHIZARD: Simulating Multi-Particle Processes at LHC and ILC}",
    eprint = "0708.4233",
    archivePrefix = "arXiv",
    primaryClass = "hep-ph",
    reportNumber = "DESY-11-126, EDINBURGH-2010-36, FR-PHENO-2010-037, SI-HEP-2010-18",
    doi = "10.1140/epjc/s10052-011-1742-y",
    journal = "Eur. Phys. J. C",
    volume = "71",
    pages = "1742",
    year = "2011"
}

@article{Alwall_2014,
    author = "Alwall, J. and Frederix, R. and Frixione, S. and Hirschi, V. and Maltoni, F. and Mattelaer, O. and Shao, H. -S. and Stelzer, T. and Torrielli, P. and Zaro, M.",
    title = "{The automated computation of tree-level and next-to-leading order differential cross sections, and their matching to parton shower simulations}",
    eprint = "1405.0301",
    archivePrefix = "arXiv",
    primaryClass = "hep-ph",
    reportNumber = "CERN-PH-TH-2014-064, CP3-14-18, LPN14-066, MCNET-14-09, ZU-TH-14-14",
    doi = "10.1007/JHEP07(2014)079",
    journal = "JHEP",
    volume = "07",
    pages = "079",
    year = "2014"
}

@software{b2luigi2025,
  author       = {Alexander Heidelbach and
                  Michael Eliachevitch and
                  Nils Braun and
                  Jonas Eppelt and
                  Giacomo De Pietro and
                  Marcel and
                  Cleora Völker and
                  anselm-baur and
                  Felix Metzner and
                  Moritz Bauer and
                  Maximilian Welsch and
                  Matthias Schnepf and
                  Tristan Fillinger and
                  Kilian Lieret and
                  Patrick Ecker and
                  Sviatoslav Bilokin},
  title        = {belle2/b2luigi: v1.2.2},
  month        = apr,
  year         = 2025,
  publisher    = {Zenodo},
  version      = {v1.2.2},
  doi          = {10.5281/zenodo.15229241},
  url          = {https://doi.org/10.5281/zenodo.15229241},
  swhid        = {swh:1:dir:16e2f7296f66c44b6aea794d873c8b9078c949bd
                   ;origin=https://doi.org/10.5281/zenodo.10853220;vi
                   sit=swh:1:snp:8715e0b28a60f6daa97220995630c8fbeefc
                   ae34;anchor=swh:1:rel:abeb7fb912e284089065560f78ed
                   f58c191ba65a;path=belle2-b2luigi-9b8b17a
                  },
}

@misc{luigi2025,
  author = {{Spotify}},
  title = {luigi},
  year = {2025},
  publisher = {GitHub},
  journal = {GitHub repository},
  howpublished = {\url{https://github.com/spotify/luigi}},
}

@article{CooperHarris:2025lqd,
    author = "Cooper Harris, Cameron and Desai, Aman",
    title = "{FLARE: FCCee b2Luigi Automated Reconstruction and Event processing}",
    eprint = "2506.16094",
    archivePrefix = "arXiv",
    primaryClass = "hep-ph",
    reportNumber = "ADP-25-23/T1285",
    doi = "10.1016/j.cpc.2026.110062",
    journal = "Comput. Phys. Commun.",
    volume = "322",
    pages = "110062",
    year = "2026"
}

@article{Sjostrand:2006za,
    author = "Sjostrand, Torbjorn and Mrenna, Stephen and Skands, Peter Z.",
    title = "{PYTHIA 6.4 Physics and Manual}",
    eprint = "hep-ph/0603175",
    archivePrefix = "arXiv",
    reportNumber = "FERMILAB-PUB-06-052-CD-T, LU-TP-06-13",
    doi = "10.1088/1126-6708/2006/05/026",
    journal = "JHEP",
    volume = "05",
    pages = "026",
    year = "2006"
}

@article{deFavereau:2013fsa,
    author = "de Favereau, J. and Delaere, C. and Demin, P. and Giammanco, A. and Lema\^\i{}tre, V. and Mertens, A. and Selvaggi, M.",
    collaboration = "DELPHES 3",
    title = "{DELPHES 3, A modular framework for fast simulation of a generic collider experiment}",
    eprint = "1307.6346",
    archivePrefix = "arXiv",
    primaryClass = "hep-ex",
    doi = "10.1007/JHEP02(2014)057",
    journal = "JHEP",
    volume = "02",
    pages = "057",
    year = "2014"
}

@article{Sjostrand:2014zea,
    author = {Sj\"ostrand, Torbj\"orn and Ask, Stefan and Christiansen, Jesper R. and Corke, Richard and Desai, Nishita and Ilten, Philip and Mrenna, Stephen and Prestel, Stefan and Rasmussen, Christine O. and Skands, Peter Z.},
    title = "{An introduction to PYTHIA 8.2}",
    eprint = "1410.3012",
    archivePrefix = "arXiv",
    primaryClass = "hep-ph",
    reportNumber = "LU-TP-14-36, MCNET-14-22, CERN-PH-TH-2014-190, FERMILAB-PUB-14-316-CD, DESY-14-178, SLAC-PUB-16122",
    doi = "10.1016/j.cpc.2015.01.024",
    journal = "Comput. Phys. Commun.",
    volume = "191",
    pages = "159--177",
    year = "2015"
}

@article{FCC:2025lpp,
    author = "Benedikt, M. and others",
    collaboration = "FCC",
    title = "{Future Circular Collider Feasibility Study Report: Volume 1, Physics, Experiments, Detectors}",
    eprint = "2505.00272",
    archivePrefix = "arXiv",
    primaryClass = "hep-ex",
    reportNumber = "CERN-FCC-PHYS-2025-0002",
    doi = "10.1140/epjc/s10052-025-15077-x",
    journal = "Eur. Phys. J. C",
    volume = "85",
    number = "12",
    pages = "1468",
    year = "2025",
    note = "[Erratum: Eur.Phys.J.C 86, 844 (2026)]"
}

@article{Key4hep:2023nmr,
    author = "Sailer, Andre and others",
    collaboration = "Key4hep",
    title = "{The Key4hep software stack: Beyond Future Higgs factories}",
    eprint = "2312.08151",
    archivePrefix = "arXiv",
    primaryClass = "hep-ex",
    doi = "10.1088/1742-6596/3206/1/012043",
    journal = "J. Phys. Conf. Ser.",
    volume = "3206",
    number = "1",
    pages = "012043",
    year = "2026"
}

@software{helsens_2025_15528870,
  author       = {Helsens, Clement and
                  Perez, Emmanuel and
                  Selvaggi, Michele and
                  Volkl, Valentin and
                  Forthomme, Laurent and
                  Munch Torndal, Julie},
  title        = {HEP-FCC/FCCAnalyses: v0.11.0},
  month        = may,
  year         = 2025,
  publisher    = {Zenodo},
  version      = {v0.11.0},
  doi          = {10.5281/zenodo.15528870},
  url          = {https://doi.org/10.5281/zenodo.15528870},
  swhid        = {swh:1:dir:7ed34e7a349b082dccf4197d7e5ab6679f729048
                   ;origin=https://doi.org/10.5281/zenodo.4767810;vis
                   it=swh:1:snp:8c0db7440ec53d2d63b5f5a811a002003e4ac
                   a54;anchor=swh:1:rel:f409bca09af6c3f5d2bfb1f14b1e2
                   9239e8cbaf5;path=HEP-FCC-FCCAnalyses-21d43d7
                  },
}

@article{condor-practice,
    author = "Thain, Douglas and Tannenbaum, Todd and Livny, Miron",
    title = "{Distributed computing in practice: the Condor experience}",
    doi = "10.1002/cpe.938",
    journal = "Concurrency Comput. Pract. Exp.",
    volume = "17",
    number = "2-4",
    pages = "323--356",
    year = "2005"
}

@misc{IDM-lsf, title={Platform LSF Version 9 Release 1.3}, url={https://www.hpc.dtu.dk/lsf931/}, howpublished={IBM Platform LSF Documentation, Version 9.1.3}, author={{IBM}}, year={2014}}

@inproceedings{slurm,
    author = "Yoo, Andy B. and Jette, Morris A. and Grondona, Mark",
    editor = "Feitelson, Dror and Rudolph, Larry and Schwiegelshohn, Uwe",
    title = "{SLURM: Simple Linux Utility for Resource Management}",
    booktitle = "{Job Scheduling Strategies for Parallel Processing (JSSPP 2003)}",
    series = "Lecture Notes in Computer Science",
    volume = "2862",
    publisher = "Springer",
    pages = "44--60",
    doi = "10.1007/10968987_3",
    year = "2003"
}

@article{Blomer:2011zz,
    author = "Blomer, Jakob and Aguado Sanchez, Carlos and Buncic, Predrag and Harutyunyan, Artem",
    editor = "Lin, Simon C.",
    title = "{Distributing LHC application software and conditions databases using the CernVM file system}",
    doi = "10.1088/1742-6596/331/4/042003",
    journal = "J. Phys. Conf. Ser.",
    volume = "331",
    pages = "042003",
    year = "2011"
}

@article{Brun:1997pa,
    author = "Brun, Rene and Rademakers, Fons",
    editor = "Werlen, M. and Perret-Gallix, D.",
    title = "{ROOT {\textemdash} An object oriented data analysis framework}",
    doi = "10.1016/S0168-9002(97)00048-X",
    journal = "Nucl. Instrum. Meth. A",
    volume = "389",
    number = "1-2",
    pages = "81--86",
    year = "1997"
}

@misc{pydantic,
  author       = {Colvin, Samuel and {Pydantic contributors}},
  title        = {Pydantic: data validation for {P}ython},
  year         = {2025},
  howpublished = {\url{https://github.com/pydantic/pydantic}},
}

@article{Desai:2026lst,
    author = "Desai, Aman",
    title = "{LHEREADER: simplified conversion from Les Houches event files to ROOT format}",
    eprint = "2603.01489",
    archivePrefix = "arXiv",
    primaryClass = "hep-ph",
    doi = "10.1088/1402-4896/ae7772",
    journal = "Phys. Scripta",
    volume = "101",
    number = "23",
    pages = "231501",
    year = "2026"
}

\end{document}